\documentclass[twocolumn,prb,aps]{revtex4}
\usepackage[dvips]{graphicx}
\begin{document}
\title{Electronic thermal conductivity at high temperatures: 
Violation of the Wiedemann-Franz law in narrow band metals}


\begin{abstract} 
We study the electronic part of the thermal conductivity 
$\kappa$ of metals. We present two methods for calculating
$\kappa$, a quantum Monte-Carlo (QMC) method and a method
where the phonons but not the electrons are treated semiclassically
(SC). We compare the two methods for a model of alkali-doped C$_{60}$,
A$_3$C$_{60}$, and show that they agree well. We then mainly use 
the SC method, which is simpler and easier to interpret.  We perform 
SC calculations for Nb for large temperatures $T$ and find that 
$\kappa$ increases with $T$ as $\kappa(T)=a+bT$, where $a$ and 
$b$ are constants, consistent with a saturation of the mean free
path, $l$, and in good agreement with experiment. In contrast,
we find that for A$_3$C$_{60}$, $\kappa(T)$ decreases with $T$ for 
very large $T$. We discuss the reason for this qualitatively in the 
limit of large $T$. We give a quantum-mechanical explanation of the 
saturation of $l$ for Nb and derive the Wiedemann-Franz law 
in the limit of $T\ll W$, where $W$ is the band width. In contrast,
due to the small $W$ of A$_3$C$_{60}$, the assumption 
$T\ll W$ can be violated. We show that this leads to $\kappa(T) 
\sim T^{-3/2}$ for very large $T$ and a strong violation of the 
Wiedemann-Franz law.

\end{abstract}
\author{K. Vafayi,$^1$ M. Calandra,$^2$ and O. Gunnarsson$^1$}   
\affiliation{
$^1$Max-Planck-Institut f\"ur Festk\"orperforschung, D-70506 Stuttgart, Germany
}
\affiliation{$^2$Institut de Min$ \acute e$ralogie et de Physique des Milieux Condens$ \acute e$s, 4 place Jussieu, 75252, Paris cedex 05, France}

\maketitle

\section{Introduction}\label{sec:1}

In the Sommerfeld theory, the electronic contribution $\kappa$
to the thermal conductivity of a metal can be written as \cite{Ashcroft}
\begin{equation}\label{eq:1}
\kappa={1\over 3}v_Flc_v={\pi^2\over 3}{nk_B^2\over mv_F}lT,
\end{equation}
where $v_F$ is the Fermi velocity, $l$ the mean free path, $c_v=
\pi^2 k_B^2 n T/(mv_F^2)$             
is the specific heat per electron, $n$ is the electron density, 
$k_B$ is the Boltzmann constant, $m$ is the electron mass 
and $T$ the temperature. For large $T$ the mean free
path decreases as $l \sim 1/T$, and one might then 
expect that $\kappa$ approximately approaches a constant for large 
$T$. This is indeed found for many good metals. However, in the 
context of electrical conductivity, it has been found that for 
large $T$ the so-called parallel resistor formula describes the 
experimental resistivity quite well for many metals with a large 
resistivity.\cite{Wiesmann}  This formula corresponds to the 
assumption
\begin{equation}\label{eq:2}
l=d+{c\over T},
\end{equation}
where $d$ is a distance of the order of the separation of two
atoms and $c$ is a constant. According to this formula the mean
free path initially decreases rapidly with $T$ but then saturates
at $l \sim d$. This formula is often justified by arguing that in
a semiclassical picture, at worst, an electron is scattered at 
every atom, leading to $l \sim d$. Inserting Eq.~(\ref{eq:2}) 
in Eq.~(\ref{eq:1}) leads to 
\begin{equation}\label{eq:3}
\kappa=a+bT,
\end{equation}
where $a$ and $b$ are constants. This formula gives a fairly good
qualitative description of experimental results for many transition metal 
and transition metal compounds with a small $\kappa$. 

The theoretical justification for Eq.~(\ref{eq:2}), however, is 
very unsatisfactory. It is based on a semiclassical picture, which
is only valid for $l\gg d$, and which cannot be used to discuss what
happens for $l\sim d$. Nevertheless, in the 1970's and early 1980's 
it seemed that $l\gtrsim d$ was a universal behavior,\cite{Allen}
satisfying the Ioffe-Regel condition.\cite{Ioffe} Later experimental 
work, however, has found many examples of metals where the resistivity 
is much larger than predicted by the Ioffe-Regel condition and where 
the apparent mean free path is much shorter than the separation
of two atoms.\cite{RMP} This illustrates that the semiclassical 
explanation for Eq.~(\ref{eq:2}) is incorrect and there is a need 
for a quantum-mechanical explanation of Eq.~(\ref{eq:3}). Such a 
theory has been presented in the context of the electrical 
conductivity.\cite{Matteo} The purpose of this paper is to provide 
a quantum-mechanical justification for the thermal conductivity 
in Eq.~(\ref{eq:3}). 

To calculate the thermal conductivity we use a Quantum Monte-Carlo (QMC)
method, which can solve the models used here accurately. To interpret
the results, we also introduce a simpler semiclassical (SC) method, 
where the phonons but not the electrons are treated semiclassically. 

For many metals, the Wiedemann-Franz law 
\begin{equation}\label{eq:4}
{\kappa\over \sigma T}={\pi^2\over 3}({k_B\over e})^2,
\end{equation}
is approximately satisfied, where $\sigma$ is the electrical 
conductivity and $e$ the electron charge. We derive this law 
by making assumptions which are expected to be reasonable 
when $T$ is so large that the transport is completely incoherent
but $T$ is still much smaller than $W$, where $W$ is the band width. For 
$T\gtrsim W$, however, we find that the Wiedemann-Franz law 
is strongly violated. This may apply to alkali-doped fullerides,
A$_3$C$_{60}$,  at or somewhat above the highest temperatures that 
can be achieved experimentally.

In Sec. \ref{sec:2} we present the models and in Sec.~\ref{sec:3}
the methods used. In Sec.~\ref{sec:4} we compare the QMC and SC
methods. The results for Nb and A$_3$C$_{60}$ in the SC method
are given in Sec.~\ref{sec:6}. These results are discussed 
qualitatively in Sec.~\ref{sec:5}.

\section{Models}\label{sec:2}

We first consider a model of Nb, referred to as the transition
metal (TM) model, which is appropriate for describing transition 
metals or transition metal compounds. Each Nb atom has a 
five-fold degenerate ($N_d=5$) level. The         
hopping matrix elements are described by $t_{\mu\nu}$, where
$\nu\equiv (m,i)$ is a combined label for a orbital index $m$ and 
a site index $i$. Thus the electronic Hamiltonian is 
\begin{equation}\label{eq:2.1}
H_0^{\rm el}=\varepsilon_0\sum_{\mu\sigma}c^{\dagger}_{\mu\sigma}
c_{\mu\sigma}^{\phantom \dagger} +\sum_{\mu\nu\sigma}t_{\mu\nu} 
c^{\dagger}_{\mu\sigma}c_{\nu\sigma}^{\phantom \dagger},
\end{equation}
where $c^{\dagger}_{\mu\sigma}$ creates an electron in the state 
$\mu$ with spin $\sigma$.
We assume that the Coulomb interaction can be neglected, since the
band width of Nb is rather large. The precise form of the hopping
matrix elements has been described elsewhere.\cite{MatteoPRB}

We assume that the electron scattering is due to the electron-phonon 
coupling. For the Nb model, we assume that the phonons couple to 
hopping integrals (HI). The phonons are approximated as Einstein 
phonons, with one phonon for each coordinate direction. The frequency 
$\omega_{ph}=0.014$ eV was set equal to the average phonon frequency 
of Nb metal.\cite{Wolf} Due to the vibrations of the atoms the hopping 
matrix elements are modulated, leading to an electron-phonon 
coupling.\cite{MatteoPRB} 

We next introduce a model of alkali-doped C$_{60}$, A$_3$C$_{60}$,
referred to as the C$_{60}$ model. We use a model including the partly 
occupied three-fold degenerate $t_{1u}$ orbital on each C$_{60}$ and 
the hopping matrix elements connecting these orbitals. This results 
in a Hamiltonian of the same general form as in Eq.~(\ref{eq:2.1})
but with a different lattice structure and different orbital degeneracy
($N_d=3$).  The hopping integrals are obtained from a tight-binding
description.\cite{Book,MatteoPRB} For simplicity, the Coulomb interaction 
is neglected, although this may be a questionable approximation. 
The main electron-phonon coupling 
is due to the intramolecular five-fold degenerate phonons of H$_g$ 
symmetry, which have an on-site Jahn-Teller coupling to the $t_{1u}$
levels of a C$_{60}$ molecule.\cite{Book,MatteoPRB} We refer to
this as a level energy (LE) coupling. For the TM and C$_{60}$ models 
we define the dimensionless coupling $\lambda$ so that the real
part of the electron-phonon part of the electron self-energy is 
given by
\begin{equation}\label{eq:2.6}
{\rm Re} \Sigma_{\rm ep}(\omega)=-\lambda \omega,
\end{equation}
in the weak-coupling limit and for small $\omega$.

To introduce current operators we follow Mahan.\cite{Mahan}
The particle current is obtained from particle conservation                  
\begin{equation}\label{eq:2.2}
{\bf  j}={i\over \hbar}[H,{\bf P}]=-{i\over \hbar}\sum_{\mu\nu\sigma}
({\bf R}_{\mu}-{\bf R}_{\nu})t_{\mu\nu}c_{\mu \sigma}^{\dagger}
c_{\nu \sigma},
\end{equation}
where $H$ is the full Hamiltonian and ${\bf P}$ is the polarization 
operator
\begin{equation}\label{eq:2.3}
P=\sum_i{\bf R}_in_i.
\end{equation}
Here ${\bf R}_{\nu}={\bf R}_i$ is the position of the $i$th atom
($\nu=(m,i)$).
In a similar way the energy current is obtained from energy 
conservation
\begin{eqnarray}\label{eq:2.4}
&&{\bf j}_E={i\over \hbar}\sum_{i}{\bf R}_i[H,h_i]    \\
&&=-{i\over 2\hbar}\sum_{\mu \gamma \nu\sigma}t_{\mu\gamma}t_{\gamma \nu}
({\bf R}_{\mu}-{\bf R}_{\nu})c_{\mu \sigma}^{\dagger}c_{\nu \sigma}
^{\phantom \dagger}, \nonumber 
\end{eqnarray}
where $h_i$ is a Hamiltonian of the $i$th atomic site, defined
in such a way that the hopping terms between two sites are
split equally between these two sites and $H=\sum_i h_i$.  
Following Ref. \onlinecite{Mahan}, we also introduce a heat current
\begin{equation}\label{eq:2.5}
{\bf j}_Q={\bf j}_E-\mu {\bf j},
\end{equation}
where $\mu$ is the chemical potential.

\section{Methods}\label{sec:3}

The conductivity is calculated using a Kubo formalism.
We assume an isotropic system and define a heat current 
- heat current correlation function\cite{Mahan}      
\begin{equation}\label{eq:3.1}
\pi^{(22)}(i\omega_n)=-{1\over 3N\Omega}\int_0^{\beta \hbar}d\tau e^{i\omega_n \tau}
\langle T_{\tau} {\bf  j}_Q(\tau)\cdot {\bf  j}_Q(0)\rangle,
\end{equation}
where $N$ is the number of atoms, $\Omega$ is the volume per atom, 
$\beta=1/(k_BT)$, $T_{\tau}$ is a 
time-ordering operator and $\omega_n$ is a Matsubara frequency. 
In a similar way we define a particle current - heat current 
correlation function $\pi^{(12)}$ and a particle current -
particle current correlation function $\pi^{(11)}$. These 
correlation functions are analytically continued, giving 
retarded response functions $\pi_{\rm ret}^{nm}(\omega)$.
Then the electrical conductivity is given by\cite{Mahan}
\begin{equation}\label{eq:3.2} 
\sigma(\omega)=-e^2 {{\rm Im} \pi^{(11)}_{\rm ret}
(\omega)\over \hbar \omega}.
\end{equation}
The heat conductivity is\cite{Mahan}
\begin{equation}\label{eq:3.3}
\kappa(\omega)=- {1\over T \hbar \omega}\lbrace {\rm Im} \pi^{(22)}_{\rm ret}
(\omega)- {\lbrack {\rm Im} \pi^{(12)}_{\rm ret}(\omega)\rbrack^2
\over {\rm Im} \pi^{(11)}_{\rm ret}(\omega)} \rbrace.
\end{equation}

We use two different methods for performing calculations, a 
determinantal quantum Monte-Carlo (QMC) method \cite{Scalapino} 
and a semiclassical (SC) method. The QMC method can be used to 
obtain properties very accurately. To interprete these results 
we use a SC method where the phonons but not the electrons are 
treated semiclassically. 

Due to the lack of a repulsive Coulomb interaction in the 
models treated here, the QMC method has no so-called sign problem. 
The statistical errors in the QMC calculation of response 
functions for imaginary times can then be made arbitrarily 
small by improving the sampling. To analytically continue these 
response functions to the real frequency axis we use a maximum 
entropy method.\cite{Jarrell} 

The Hamiltonians treated here can be written in the form
\begin{eqnarray}\label{eq:3.4}
H&&=H^{\rm el}_0+\sum_{\mu \nu n \sigma}g_{\mu\nu n}
c_{\mu \sigma}^{\dagger}c_{\nu\sigma}^{\phantom \dagger}q_n
+{1\over 2}\sum_n(p_n^2+\omega_{\rm ph}^2 q_n^2) \nonumber \\ 
&&\equiv H^{\rm el}({\bf q})+H_B,
\end{eqnarray}
where $g_{\mu\nu n}$ is a coupling constant, $q_n$ is a phonon 
coordinate, $p_n$ a phonon momentum and $\omega_{\rm ph}$ is the
frequency of the phonons. In the QMC calculation 
the starting point is the partition function
\begin{equation}\label{eq:3.5}
Z={\rm Tr}\int d{\bf q}\langle {\bf q}| e^{-\beta H}|{\bf q}\rangle
\end{equation}
where Tr is a trace over all electronic states and the integral over 
${\bf q}\equiv (q_1, q_2, ...)$ results in a trace over ${\bf q}$. 
Since $H^{\rm el}({\bf q})$ and $H_B$ do not commute, we use a Trotter 
decomposition and define $\Delta \tau= \beta \hbar/L$, where $L$ is 
some integer. We write exp$(-\beta H)$ as $\Pi_{i=1}^L{\rm exp}
(-\Delta \tau H_{el}({\bf q})){\rm exp}(-\Delta \tau H_B)$. 
By inserting complete sets of phonon states 
between each factor, one obtains\cite{Scalapino,Hirsch}

\begin{eqnarray}\label{eq:3.6}
&&Z={\rm Tr}\int(\Pi_{i=1}^L d{\bf q}_i)e^{-\Delta \tau H_{el}({\bf q}_1)}
\langle {\bf q}_1|e^{-\Delta \tau H_B}|{\bf q}_L \rangle \nonumber \\ 
&&\times .... e^{-\Delta \tau H_{el}({\bf q}_2)}
\langle {\bf q}_2|e^{-\Delta \tau H_B}|{\bf q}_1 \rangle
\end{eqnarray}
The phonon coordinates ${\bf q}_i\equiv \lbrace q_{n,i}\rbrace 
\equiv \lbrace q_n(\tau_i) \rbrace$ are sampled by using a 
Monte-Carlo approach with the integrand as a weight factor. Here 
$\tau_i=i\Delta \tau$ is an imaginary time. Correlation functions 
$\langle X(\tau_i)
X(\tau_j)\rangle$ are obtained in a similar way by inserting 
the operators at positions corresponding to $\tau_i$ and 
$\tau_j$.\cite{Scalapino,Hirsch} For the case of the HI coupling 
there are technical complications in terms of updating the relevant 
quantities after each Monte-Carlo step, which are discussed 
elsewhere.\cite{MatteoPRB} 

In the SC treatment of the phonons,\cite{MatteoPRB} 
we introduce a super cell with periodic boundary conditions. 
Each phonon coordinate is given a random displacement according 
to a Gaussian distribution centered at zero and with the width
\begin{equation}\label{eq:me2}
\langle x^2 \rangle = {\hbar \over M\omega_{ph}}[n_B(T)+0.5]
\end{equation}
where
\begin{equation}\label{eq:me2a}
n_B(T)={1\over e^{\hbar \omega_{ph}/(k_B T)}-1},
\end{equation}
is the occupation of the phonon mode. For fixed phonon coordinates,
the resulting Hamiltonian is a one-particle Hamiltonian,
which can easily be diagonalized. The response functions  
$\pi^{(nm)}$ can then be expressed in terms of the eigenstates 
$|l\rangle$ and eigenvalues $\varepsilon_l$ of this Hamiltonian as 
\begin{eqnarray}\label{eq:me3}
&&{\rm Im}\pi^{(mn)}_{xx}(\omega) =-{2\pi \hbar\over N\Omega} 
\sum_{ll^{'}}
\langle l| (j_m)_x| l^{'}\rangle 
\langle l^{'}| (j_n)_x| l\rangle  \nonumber \\ 
&& \times (f_l-f_{l^{'}})\delta(\hbar \omega +\varepsilon_{l}
-\varepsilon_l^{'}),
\end{eqnarray}
where $j_1=j$, $j_2=j_Q$ and $f_l$ is the Fermi function for 
the energy $\varepsilon_l$.    
We have here for simplicity considered the $xx$ component.
The result is averaged over different configurations
of random displacements of the atoms.

The electron-phonon interaction leads to a renormalization 
of the phonon frequency. This renormalization acts back on 
the electronic properties. This is included in the QMC 
approach above, but not in the SC method. To be able to 
compare more directly with this SC method, we also perform 
QMC calculations where the electron-phonon interaction 
is neglected when calculating the weight functions entering 
the sampling in Eq.~(\ref{eq:3.6}). In all other parts of 
the QMC calculation the electron-phonon interaction is fully 
taken into account. The result is that the phonon coordinates 
are sampled as if the phonons were free phonons. In {\it this} 
QMC calculation, the phonons are then not renormalized, and 
the corresponding influence on the electronic properties is 
absent. 

\section{Comparison of QMC and semiclassical methods}\label{sec:4}

\begin{figure}
{\rotatebox{-90}{\resizebox{6.0cm}{!}{\includegraphics {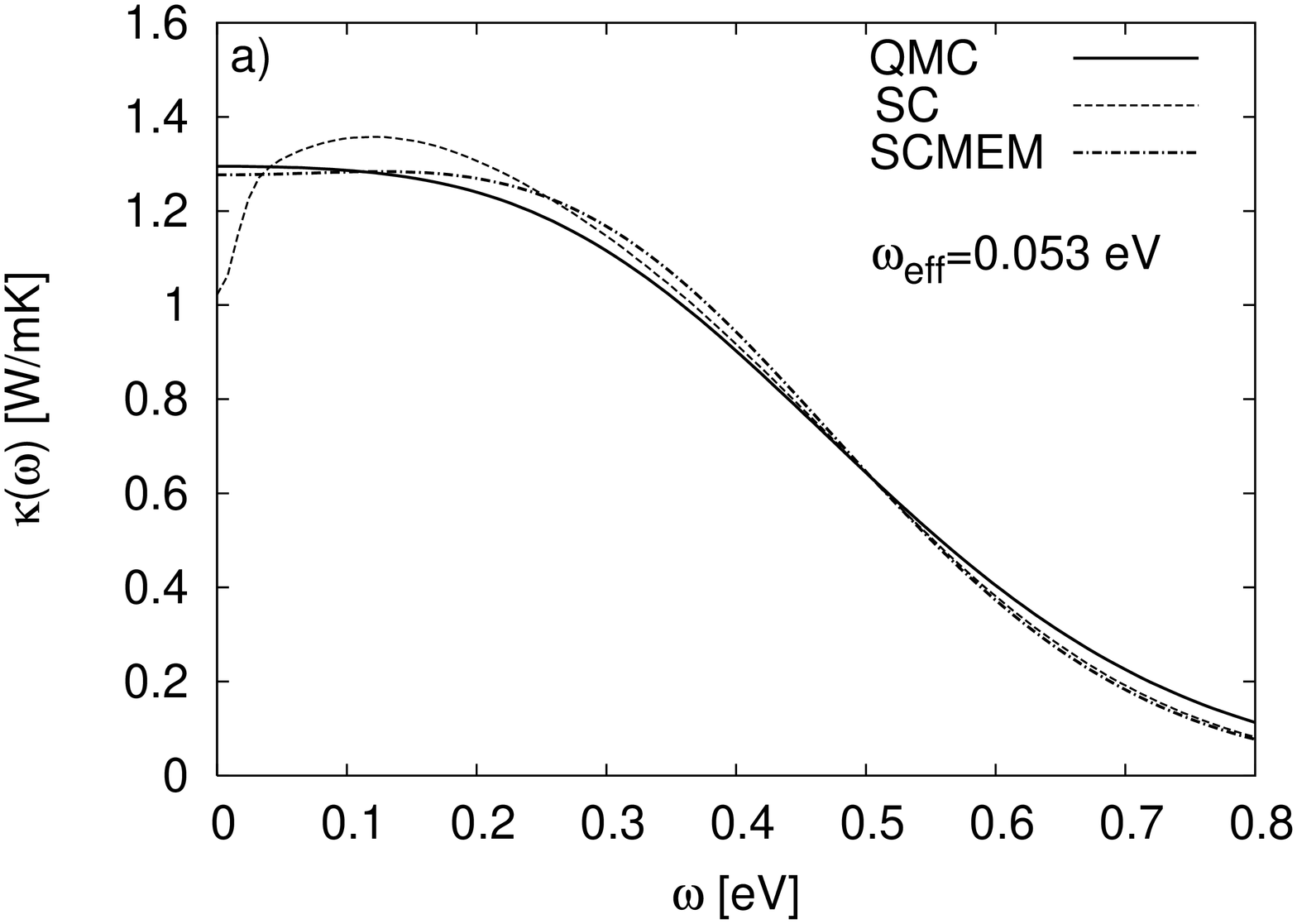}}}}
{\rotatebox{-90}{\resizebox{6.0cm}{!}{\includegraphics {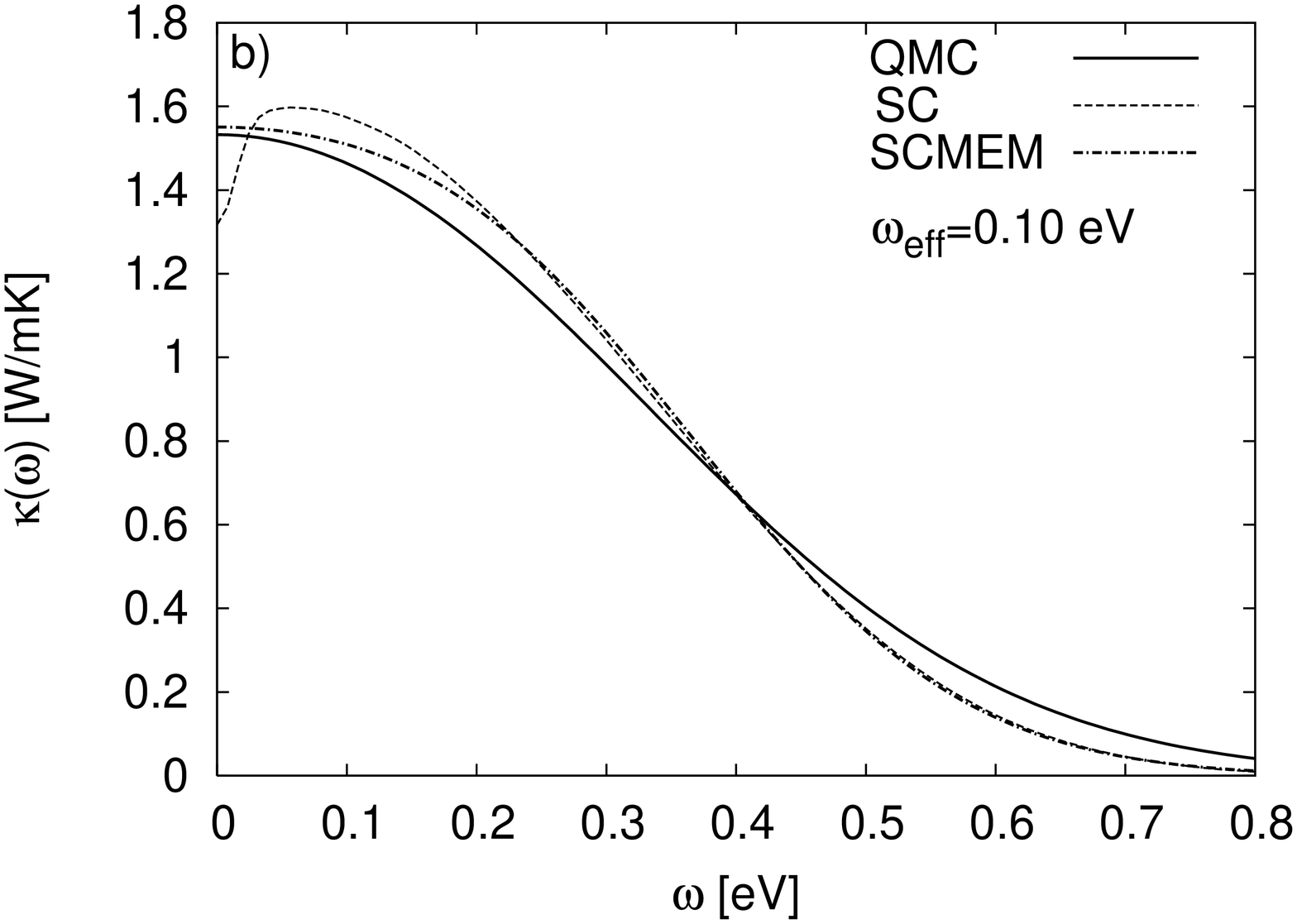}}}}
\caption{\label{fig:1}a) Thermal conductivity $\kappa(\omega)$ of 
A$_3$C$_{60}$ as a function of $\omega$ according to QMC (full line),
SC (dashed line) and SCMEM (dashed-dotted line). b) The same as in a)
but with the renormalization of the phonon frequency suppressed. The 
parameters are $\omega_{\rm ph}=0.1$ eV, $\lambda=0.5$ and $T=0.1$ eV. 
In a) $\omega_{ph}$ is renormalized to $\omega_{\rm eff}=0.053$ eV. 
The model of A$_3$C$_{60}$ described in Sec. \ref{sec:2} was used. 
The calculations were done for a cluster with $3\times 4\times 4$
sites and averaged over 20000 different random configurations.
A Gaussian broadening with FWHM 0.03 eV was used. The figure 
illustrates that the QMC and SCMEM calculations agree very well 
and that the downturn at $\omega=0$ in the SC results is lost in 
the SCMEM results due to the MEM procedure.
}
\end{figure}

To test the accuracy of the SC method, we compare the thermal 
conductivity of the C$_{60}$ model according to the QMC and SC 
method. As discussed in Sec. \ref{sec:3}, in the QMC calculation 
the phonon frequency is renormalized by the electron-phonon 
interaction, while this is not the case in the SC calculation. 
To compare the two methods we use two different approaches. 
In one approach we suppress the phonon renormalization 
in the QMC calculation, as discussed in Sec. \ref{sec:3}. In 
a second approach we calculate an effective phonon frequency,
$\omega_{\rm eff}$, in the QMC program and then put in this 
frequency by hand in the SC calculation, keeping the coupling
constants fixed. To obtain the effective
phonon frequency we calculate the phonon Green's function 
$D(\tau)$ for imaginary times $\tau$. We compare this with
the phonon Green's function, $D_0(\tau,\omega_{\rm eff})$, 
for noninteracting phonons with the frequency $\omega_{\rm eff}$. 
We minimize   
\begin{equation}\label{eq:c1}
\sum_i \lbrack D(\tau_i)-D_0(\tau_i,\omega_{\rm eff}) \rbrack^2
\end{equation}
with respect to $\omega_{\rm eff}$, where $\tau_i$ correspond to
the discrete values of $\tau$ used in the QMC calculation.  
The electron-phonon coupling constants are kept unchanged.

In the SC method $\kappa(\omega)$ is calculated directly for 
real frequencies, while in the QMC method we calculate the 
response functions $\pi^{nm}(\tau)$ for imaginary times and then 
use a maximum entropy method (MEM) to perform the analytical
continuation. To make the QMC and SC methods more comparable,
we have therefore in the SC method also made a transformation 
of the data to imaginary times for each random configuration, 
which is an accurate and well-behaved transformation. We have then  
transformed the data back to real frequencies, using MEM. These 
results, referred to as SCMEM results, then contain errors 
introduced by the analytical continuation, and are in this sense 
more comparable to the QMC results.  

Fig. \ref{fig:1} shows results for $\kappa(\omega)$ of the C$_{60}$
model for $T=0.1$ eV, $\omega_{ph}=0.1$ eV and $\lambda=0.5$. In the QMC 
calculation the phonon frequency is renormalized to $\omega_{\rm eff}
=0.053$ eV.  In Fig. \ref{fig:1}a the phonon renormalization is taken 
into account and in Fig. \ref{fig:1}b it is suppressed. It is interesting 
that the QMC and SCMEM data agree quite well, both with and without 
renormalization of the phonon frequency. This suggests that the SC 
method is rather accurate. We also notice that the renormalization
of the phonon frequency leads to an effectively stronger electron-phonon
coupling and therefore a smaller $\kappa(\omega=0)$. This follows 
since the coupling goes as the inverse phonon frequency if the
coupling constants are kept fixed, as is done here.   

The SC results show a downturn for small $\omega$. This may be 
the beginning of an Anderson localization.\cite{Ramakrishnan} 
In the SC method, the phonons introduce a static (diagonal) disorder
which leads to an  Anderson localization when the disorder becomes
sufficiently strong. This is not expected to happen in the QMC 
calculation, since this calculation takes into account that the 
scattering is inelastic and phase information is lost.\cite{Ramakrishnan}
We note, however, that if such a downturn actually would occur in 
the QMC spectrum, it would be lost in the analytical continuation 
to real frequencies. This can be seen by comparing the SC and SCMEM 
curves, since the downturn in the SC results is lost in the SCMEM
results, due to the MEM procedure. The reason is that the downturn
happens on such a small energy scale that it is not detected by 
the MEM.

Since there is a rather good agreement between the QMC and SC methods,
we use the SC method in the rest of the paper. The reason is that
the SC metod is simpler, and it is easier to interpret the results.

\section{Results using the semiclassical method}\label{sec:6}

\begin{figure}[t]
\centerline{
{\rotatebox{-90}{\resizebox{6.0cm}{!}{\includegraphics {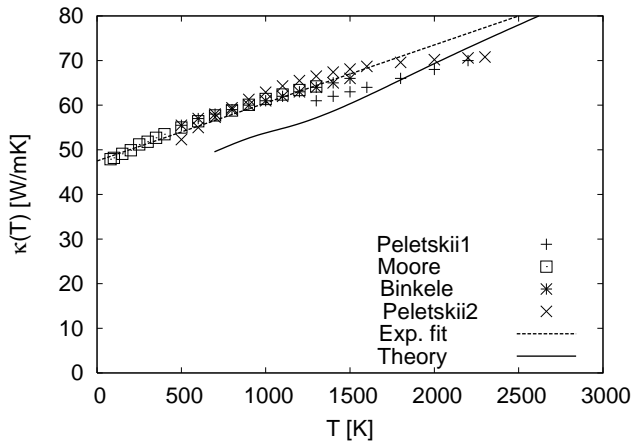}}}}}
\caption{\label{fig:2}Thermal conductivity of Nb metal according to 
the SC method (full line) and from experiments (``Peletskii1''\cite{75Pe1},
``Moore''\cite{80Mo1}, ``Binkele''\cite{86Bi1}, ``Peletskii2''\cite{87Pe1})
together with the line $\kappa(T)=47.5+0.013T$ adjusted to the
experimental results. The calculated result was obtained for a 
$10\times 10 \times 10$ lattice using a Gaussian broadening with the
FWHM of 0.06 eV. The results was averaged over 10 different 
random displacements of the atoms. The model of Nb described in Sec. 
\ref{sec:2} was used. The figure shows that theory and experiment 
agree rather well and that $\kappa(T)$ increases approximately linearly 
with $T$. 
}
\end{figure}

We have calculated the thermal conductivity of Nb using the SC method.
We used a cluster of $10\times 10 \times 10$ atoms and averaged over 
10 different distributions of thermally displaced atoms. The model 
described in Sec. \ref{sec:2} was used. The results are compared 
with experimental results \cite{75Pe1,80Mo1,86Bi1,87Pe1} in Fig. 
\ref{fig:2}. Theory and experiment agree rather well. The figure 
also shows the curve $\kappa(T)=(47.5+0.013T)$ W/mK fitted to the 
experimental results. 
This linear dependence on $T$ is in qualitative agreement with the 
result expected from the discussion in Sec. \ref{sec:5} below. 
$\kappa(T)$ increases with $T$, which within the framework of the 
discussion in the introduction would correspond to the saturation of 
the apparent mean free path [see Eq.~(\ref{eq:3})].

\begin{figure}[t]
\centerline{
{\rotatebox{-90}{\resizebox{6.0cm}{!}{\includegraphics {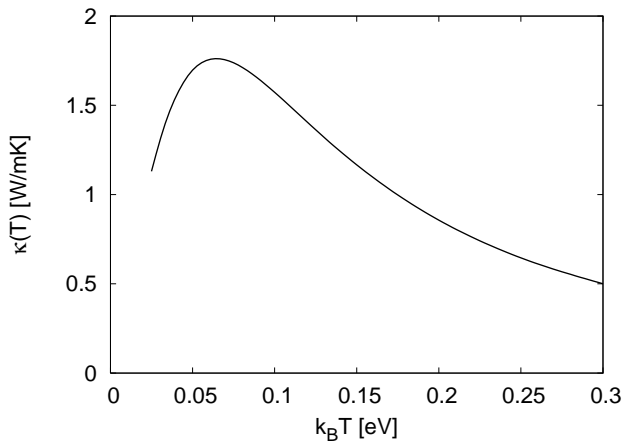}}}}}
\caption{\label{fig:3}Thermal conductivity of the C$_{60}$ model 
according to the SC method for $\lambda=0.5$ and $\omega_{ph}=0.1$ eV. 
400 random configurations for a  $8\times 8 \times 8$ cluster were 
generated, and a Gaussian broadening of 0.03 eV FWHM was used. 
}
\end{figure}

The thermal conductivity of A$_3$C$_{60}$ was calculated using the SC 
method for a cluster of $8 \times 8 \times 8$ molecules and averaging
over 400 different distributions of thermally distorted molecules. 
The parameters were $\lambda=0.5$ and $\omega_{ph}=0.1$ eV, where 
$\omega_{ph}=0.1$ eV is an approximate average renormalized phonon
frequency for A$_3$C$_{60}$. The results 
are shown in Fig. \ref{fig:3}. The results are influenced by a downturn 
for small $\omega$ (see Fig. \ref{fig:1}), which depends on the amount 
of broadening used. For the temperatures shown in Fig.~\ref{fig:3} 
($k_BT \le 0.3$ eV), however, this does  not strongly influence the results.
$\kappa(T)$ initially increases rapidly with $T$, but then reached a 
maximum followed by a rapid drop with $T$. This behavior differs 
qualitatively from the results for Nb. This difference is discussed 
extensively in the next section.

\section{Qualitative discussion in semiclassical formalism}\label{sec:5}

In this section we use the semiclassical formalism to discuss 
qualitatively both $T\ll W$ and $T \sim W$, where $W$ is the band width.
For this purpose we first derive a relation between matrix elements 
of the electrical and heat current operators. In the spirit of the 
semiclassical approximation, we assume noninteracting electrons.

Let $|l\rangle$ be a one-particle eigenstate with the energy 
$\varepsilon_l$. Using the definition of ${\bf j}_E$ in 
Eq.~(\ref{eq:2.4})
\begin{eqnarray}\label{eq:5.1}
&&\langle l |{\bf j}_E |l^{'}\rangle={i\over \hbar}(\varepsilon_l-
\varepsilon_{l^{'}})\sum_i {\bf R}_i\langle l |h_i|l^{'}\rangle \nonumber  \\ 
&&={i\over 2\hbar}(\varepsilon_l-\varepsilon_{l^{'}})(\varepsilon_l+\varepsilon_{l^{'}})
\sum_i{\bf R}_i\langle l |n_i|l^{'}\rangle \\
&&={1\over 2}(\varepsilon_l+\varepsilon_{l^{'}})
\langle l |{\bf j} |l^{'}\rangle \nonumber
\end{eqnarray}
It then follows that
\begin{equation}\label{eq:5.2}
\langle l |{\bf j}_Q |l^{'}\rangle=({\varepsilon_l+
\varepsilon_{l^{'}}\over 2}-\mu)\langle l |{\bf j} |l^{'}\rangle.
\end{equation}

We now consider values of $T$ which are sufficiently large to give 
a large thermal disorder. As a result there are matrix elements of 
the current operators between essentially all states. We therefore
make the assumption that all the matrix elements of the particle 
current operator between different eigenstates of the Hamiltonian 
have the same value $j_{av}$. This assumption has been checked 
extensively and found to lead to accurate results for $\sigma(\omega)$
for large $T$.\cite{MatteoPRB} It is important to make this
assumption for the particle current operator, since it would 
obviously not be true for the heat current operator, as can
be seen from Eq.~(\ref{eq:5.2}), due to the factor $[(\varepsilon+
\varepsilon^{'})/2-\mu)]$. 

We first consider the case when $T\ll W$. We can assume that 
the density of states per orbital and spin, $N(\varepsilon)$, 
is a constant $N(\mu)$, since states close to $\varepsilon=\mu$ 
mainly influence Im $\pi^{nm}(\omega)$ for small $\omega$ and 
$T \ll W$.  This gives
\begin{eqnarray}\label{eq:5.3}
&&{\rm Im}\pi^{(22)}_{xx}(\omega)=-{2\pi \hbar\over N\Omega}\lbrack NN_dN(\mu)
j_{av}\rbrack^2   \\
&&\times \int d\varepsilon d\varepsilon^{'}({\varepsilon +
\varepsilon^{'}\over 2}-\mu)^2\lbrack f(\varepsilon)-f(\varepsilon^{'})
\rbrack \delta(\hbar \omega+\varepsilon-\varepsilon^{'}). \nonumber
\end{eqnarray}
We now consider the limit $\omega \ll T$. We also use the assumption
$T\ll W$ to extend the integrations in Eq.~({\ref{eq:5.3}) to infinity. 
To leading order in $\omega$ we then obtain
\begin{equation}\label{eq:5.4}
{\rm Im}\pi^{(22)}_{xx}(\omega) 
=-{2\pi^3 (k_BT)^2 \over 3\Omega}N\lbrack N_dN(\mu)
j_{av}\rbrack^2 
\hbar^2 \omega. 
\end{equation}
Similar approximations lead to 
\begin{equation}\label{eq:5.5}
{\rm Im} \pi^{(12)}_{xx}(\omega)=0
\end{equation}
and
\begin{equation}\label{eq:5.6}
{\rm Im}\pi^{(11)}_{xx}(\omega)=-{2\pi\over \Omega} N
\lbrack N_dN(\mu)j_{av}\rbrack^2\hbar^2 \omega.
\end{equation}
From Eqs.~(\ref{eq:3.2},\ref{eq:3.3}) we then obtain 
\begin{equation}\label{eq:5.7}
{\kappa(0)\over \sigma(0)T}={\pi^2\over 3}({k_B\over e})^2,
\end{equation}
which is the Wiedemann-Franz law.

To obtain further understanding, we make some more explicit 
estimates. We first derive a sum rule for the particle current 
matrix elements.
\begin{equation}\label{eq:5.8}
\sum_{ll^{'} \alpha}|j_{ll^{'},\alpha}|^2= 
({ d\over \hbar})^2 \sum_{\mu\nu}|t_{\mu\nu}|^2=
{NN_dd^2\over \hbar^2}\langle \varepsilon^2\rangle,
\end{equation}
where $\alpha$ labels a coordinate, $d$ is the separation 
of two atoms and $\langle \varepsilon^2\rangle$ is the second 
moment of the density of states per site, orbital and spin. 
On the left hand side, the states $|l\rangle$ refer to extended 
eigenstates. In the middle expression,
we have made a unitary transformation to localized basis states and
used Eq.~(\ref{eq:2.2}), assuming nearest neighbor hopping only.           
Since the number of matrix elements on the left hand side of 
Eq.~(\ref{eq:5.8}) is $3(NN_d)^2$, we obtain
\begin{equation}\label{eq:5.8a}
|j_{av}|^2={d^2\over 3NN_d \hbar^2}\langle\varepsilon^2\rangle.
\end{equation}
Typically, $\langle\varepsilon^2\rangle N(\mu)^2\approx 
0.1$ for a system close to half-filling.\cite{MatteoPRB} 
Inserting this in Eqs.~(\ref{eq:3.3}, \ref{eq:5.4}) we obtain
\begin{equation}\label{eq:5.9}
\kappa={0.2\pi^3\over 9}{d^2\over \hbar \Omega}N_dk_B^2T=
0.9N_d{k_B^2T\over \hbar d},
\end{equation}
where we have used that $\Omega/d^2=(4/3\sqrt{3})d$, as 
appropriate for Nb metal with a bcc lattice. This gives 
the large $T$ behavior $\kappa=0.027T$ W/Km for Nb, where $T$ is in 
K. This could be compared with the linear behavior fitted to
experimental result\cite{Landolt} $\kappa=(47.5+0.013T$) W/Km (see Sec.
\ref{sec:6}), where the linear term in $T$ is about a factor of two
smaller than our simple estimate. 

Comparing the result in Eq.~(\ref{eq:5.9})
with Eq.~(\ref{eq:1}) we find that we have to assume that
\begin{equation}\label{eq:5.10}
l\approx 0.8 N_d^{1/3} d,
\end{equation}
to obtain the estimated thermal conductivity. This is essentially 
the Ioffe-Regel condition.

We now discuss the reasons for the saturation of the mean free
path and the linear increase of $\kappa$ with $T$, using the 
semiclassical treatment of the phonons. At small $T$, there are 
important intraband transitions for small wave vectors ${\bf q}$, 
which make contributions to Im $\pi^{(22)}(\omega)$ for small $\omega$. 
As $T$ is increased, the increasing thermal disorder leads to
a strong violation of momentum conservation (in the electronic 
system) and many transitions that were strongly suppressed at small
$T$ become important for large $T$. At the same time the transitions
corresponding to small $\omega$ are reduced. However, the 
sum rule in Eq.~(\ref{eq:5.8}) shows that the sum over all
particle current matrix elements squared is not reduced, as 
long as the hopping integrals are not reduced. Thus even if
we assume that the particle current matrix elements corresponding 
to low energy transitions are not larger than the average of 
the matrix elements, there is still an appreciable (and saturating)
weight of the transitions corresponding to a small energy transfer 
in Eq.~(\ref{eq:5.3}). The prefactor $\lbrack (\varepsilon+
\varepsilon^{'})/2-\mu \rbrack^2$ in Eq.~(\ref{eq:5.3}) furthermore 
gives increasingly large contributions 
as $T$ is increased and the Fermi functions are broadened. This 
leads to the increase in $\kappa(T)$ with $T$ and it corresponds to 
the increase in the specific heat per electron with $T$ entering 
in the Sommerfeld theory leading to Eq.~(\ref{eq:1}).

We next consider the case when $T\sim W$. This is uninteresting for
Nb, due to its large band width, but of relevance for A$_3$C$_{60}$ 
for which the band width is small. In A$_3$C$_{60}$ the phonons 
couple to the level energies. As a result the hopping integrals in 
Eq.~(\ref{eq:5.8}) are not changed and $j_{av}$ is unchanged as 
$T$ is increased. On the other hand, due to the fluctuations of
the level energies, the band width grows with $T$ as\cite{MatteoPRB}
\begin{equation}\label{eq:5.11}
W(T)=W(T=0)\sqrt{1+c\lambda{k_BT\over W(T=0)}},
\end{equation}
where $c$ is of the order 15. Thus there are substantial effects
on the band width already for $k_BT\sim W/10$. To describe these
effects we assume that the density of states can be written as
\begin{equation}\label{eq:5.12}
N(\varepsilon,T)={1\over \sqrt{1+T/T_0}}n({\varepsilon \over 
\sqrt{1+T/T_0}}),
\end{equation}
where $n(\varepsilon)$ is assumed to have no explicit $T$ 
dependence and $k_BT_0 \approx 0.1$ eV. We have        
\begin{eqnarray}\label{eq:5.13}
&&\sigma(\omega)={2\pi e^2 N \over \omega \Omega}(N_d j_{av})^2
\int d\varepsilon\int d\varepsilon^{'}N(\varepsilon,T)
N(\varepsilon^{'},T)  \nonumber  \\
&&\times \lbrack f(\varepsilon)-f(\varepsilon^{'})\rbrack 
\delta(\hbar \omega+\varepsilon-\varepsilon^{'}). 
\end{eqnarray}
In Ref. \onlinecite{MatteoPRB}, a qualitative discussion of the
electrical conductivity was given, neglecting finite $T$ effects
on the Fermi functions and emphasizing the difference between 
coupling to the level positions and hopping integrals. Thus 
the Fermi functions were replaced by $\Theta$-functions. If
this is done in Eq.~(\ref{eq:5.13}) and $j_{\rm av}$ is assumed 
to be $T$ independent, we obtain that $\sigma(0)
\sim 1/(1+T/T_0)$ and the integral over $\sigma(\omega)$,
entering the f-sum rule, goes as $1/\sqrt{1+T/T_0}$, in agreement
with the results in Ref. \onlinecite{MatteoPRB}. For the
treatment of the thermal conductivity, however, it is crucial 
to include the $T$ dependence of the Fermi functions. The 
$\varepsilon^{'}$ integration can trivially be performed, leading
to a factor $f(\varepsilon)-f(\varepsilon+ \hbar \omega)$. Assuming 
that $T \gg W$,  we can approximate this factor as $\hbar 
\omega/(4k_BT)$. This gives
\begin{eqnarray}\label{eq:5.14}
&& \sigma(\omega)={\pi e^2\hbar N \over 2 \Omega T\sqrt{1+T/T_0}}
(N_dj_{av})^2     \nonumber \\
&&\times \int dx n(x)
n(x+{\hbar \omega \over \sqrt{1+ T/T_0}}) \\
&& \equiv {1\over (T/T_0)\sqrt{1+T/T_0}}\tilde g({\hbar \omega \over \sqrt{1+T/T_0}}),
\nonumber
\end{eqnarray}
i.e., for very large $T$, $\sigma(0)$ decays as $T^{-3/2}$.
Similar calculations for the thermal conductivity give
\begin{eqnarray}\label{eq:5.15}
&& \kappa(\omega)={\pi \hbar N \sqrt{1+T/T_0}\over 2 T^2}
(N_dj_{av})^2 \nonumber \\
&&\times  \int dx n(x)n(x+{\hbar \omega \over \sqrt{1+T/T_0}}) 
(x+{\hbar \omega \over 2\sqrt{1+T/T_0}})^2  \nonumber  \\
&&\equiv {\sqrt{1+T/T_0} \over (T/T_0)^2}
\tilde h({\hbar \omega \over \sqrt{1+T/T_0}}), 
\end{eqnarray}
As a result, the Wiedemann-Franz law is strongly violated and $\kappa(0)/
\sigma(0)\sim {\rm const}$ instead of $T$ for very large $T$.

An important difference between $T \ll W$ and $T\gg W$ is the behavior
of the specific heat. 
For a fixed band width and a small $T$, the energy of noninteracting
electrons  increases as $T^2$ due to the smearing out of the Fermi 
function. This leads to an increase in the specific heat per 
electron, $c_v \sim T$, as shown in Eq.~(\ref{eq:1}), and results
in $\kappa/\sigma \sim T$. For $T \gtrsim W$, however, the electrons 
are already rather evenly distributed over the band width, and an 
additional increase of $T$ does not increase the total energy very 
much. The result is that $c_v$ then decreases with $T$. For such 
large $T$ the semiclassical theory is invalid, and it is not possible 
to express $\kappa$ in terms of $l$ and $c_v$. 
Nevertheless, it is suggestive that  $\kappa/\sigma \sim T$ is
violated, although the proper power of $T$ is not predicted 
by these arguments.

\begin{figure}
{\rotatebox{-90}{\resizebox{6.0cm}{!}{\includegraphics {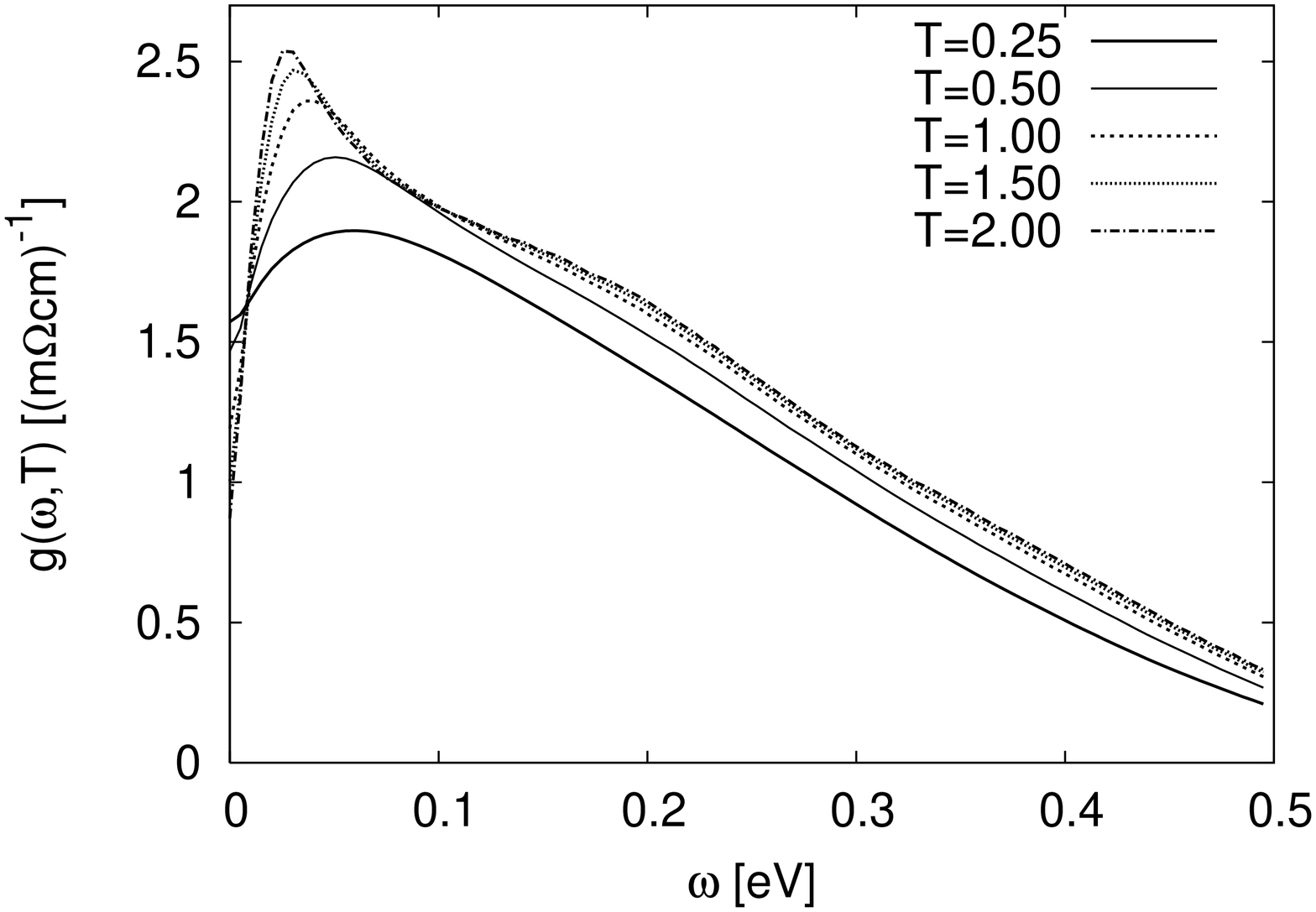}}}}
{\rotatebox{-90}{\resizebox{6.0cm}{!}{\includegraphics {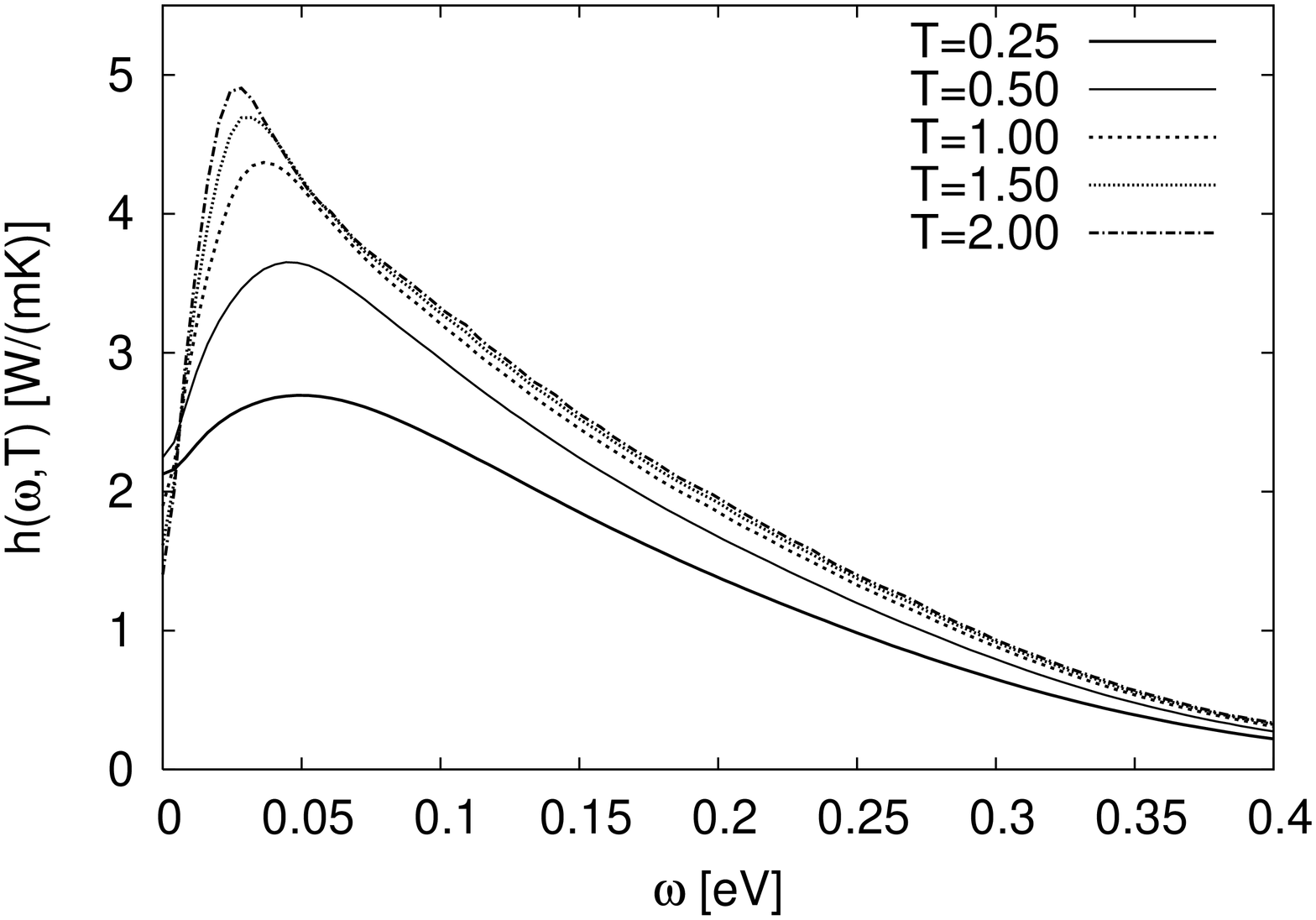}}}}
\caption{\label{fig:4}Functions $ g(\hbar \omega,T)$ and 
$ h(\hbar \omega,T)$, defined in Eq.~(\ref{eq:5.16}), as a
function of $\omega$ for different values of $k_BT=$ 0.25, 0.5, 1.0, 1.5 
and 2.0 eV calculated using the SC method. The parameters are 
$\lambda=0.5$ and $\omega_{\rm ph} =0.1$ eV. 400 random configurations
for a  $8\times 8 \times 8$ cluster were generated, and a Gaussian   
broadening of 0.03 eV FWHM was used. If the approximations 
behind Eqs.~(\ref{eq:5.14},\ref{eq:5.15}) were exact, the curves 
would fall on top of each other. 
}
\end{figure}

To test the accuracy of these arguments, we define 
\begin{equation}\label{eq:5.16}
h(\hbar \omega,T)={(T/T_0)^2\over \sqrt{1+T/T_0}}\kappa(\omega \sqrt{1+T/T_0}),
\end{equation}
and introduce an equivalent definition for $ g(\hbar \omega,T)$
in terms of $\sigma(\omega)$. These functions are shown in Fig. 
\ref{fig:4} for different values of $T$. If the approximations 
introduced  above were exact, the curves calculated for different
values of $T$ would be identical. For the values of $T$ shown in 
Fig.\ref{fig:4}, this requirement is quite well fulfilled for $g(\omega,T)$
and less well fulfilled for $h(\omega,T)$. The main source of error
is that the approximation for $f(\varepsilon)-f(\varepsilon+ \hbar \omega)$
in Eq.~(\ref{eq:5.14}) is only valid for $k_BT\gg W$, and Fig.\ref{fig:4}
shows results for $k_BT\ge W/3$. Nevertheless, varying $T$ by a factor
of almost ten only leads to a variation of the maximum values of 
$h(\omega,T)$ by less than a factor of two, providing support 
for the analysis above.

\section{Conclusions}\label{sec:8}

We have studied the electronic part $\kappa$ of the thermal 
conductivity of metals, using two different method. Accurate results
are obtained by using a determinantal Quantum Monte-Carlo (QMC)
method together with a maximum entropy method (MEM). As a much 
simpler approach we use a method where the phonons but not 
electrons are treated semiclassically (SC). We applied these
methods to a model of A$_3$C$_{60}$ (A= K, Rb) and showed 
that they give very similar results for $\kappa(\omega)$ over 
most of the energy range if the renormalization of the phonon 
frequency is treated in the same way in both methods. For very 
small values of $\omega$, however, the SC method gives a
downturn. This is probably due to an incipient Anderson localization 
transition and a defect of the SC method neglecting that the  
scattering processes are inelastic. Applying the SC method to 
a model of Nb metal, we find that the results agree well with
experiment. In particular, we find that $\kappa(T)$ increases
with $T$, consistent with a saturation of the mean free path.
In contrast, for A$_3$C$_{60}$ at very large $T$, we find a 
decrease of $\kappa(T)$ with $T$. The results are analyzed in
the SC method. To discuss very large $T$ we use an approximation 
where all matrix elements of the electric current operator are
assumed to be the equal. We can then qualitatively reproduced the 
calculated results for Nb, and give a quantum mechanical 
understanding of why the apparent mean free path saturates. 
Within this framework it is also possible to derive the 
Wiedemann-Franz law by assuming that $T \ll W$, where $W$ is
the band width. Due to the small band width of A$_3$C$_{60}$,
we instead focus on the case $T\gtrsim W$ for this system.
We then find that $\kappa(T) \sim T^{-3/2}$ indeed decreases
with $T$ and that the Wiedemann-Franz law is qualitatively 
violated.

\end{document}